\documentclass[showpacs,aps,amssymb,floatfix,prd,amsmath,preprintnumbers,nofootinbib]{revtex4}
\usepackage{graphicx}  
\usepackage{dcolumn}    
\usepackage[utf8]{inputenc}
\usepackage[T1]{fontenc}
\usepackage{bm}
\usepackage{color}
\usepackage{xcolor}
\usepackage{hyperref}
\newcommand{\be}{\begin{equation}}
	\newcommand{\ee}{\end{equation}}
\newcommand{\bea}{\begin{eqnarray}}
	\newcommand{\eea}{\end{eqnarray}}

\begin{document}

	\title{Asymptotic dynamical analysis of $f(R,T^\phi) = R+\alpha T^\phi + \beta (T^\phi)^2/2$ cosmology}
	
	\author{Joaquin Estevez-Delgado}
	\email{joaquin@fismat.umich.mx}
	\author{Roberto De Arcia} 
	\email{roberto.solis@umich.mx}
	\affiliation{Facultad de Ciencias F\'isico Matem\'aticas,
		Universidad Michoacana de San Nicol\'as de Hidalgo,
		Edificio B, Ciudad Universitaria, CP 58030,
		Morelia, Michoac\'an, M\'exico.}
	
	\author{Gabino Estevez-Delgado}
	\email{gabino.estevez@umich.mx}
	\affiliation{Facultad de Qu\'imico Farmacobiolog\'ia,
		Universidad Michoacana de San Nicol\'as de Hidalgo,
		Tzintzuntzan No.\ 173, Col.\ Matamoros, C.P.\ 58240,
		Morelia, Michoac\'an, M\'exico.}
	
	\author{Israel Quiros}
	\email{iquiros@fisica.ugto.mx}
	\affiliation{Departamento de Ingenier\'ia Civil,
		Divisi\'on de Ingenier\'ias,
		Universidad de Guanajuato, Guanajuato, M\'exico.}

	\begin{abstract}
		In this work we investigate the asymptotic cosmological dynamics of a modified gravity model based on the $f(R,T^\phi)$ theory, where $R$ denotes the Ricci scalar and $T^\phi$ is the trace of the stress-energy tensor of a scalar field. Despite the extensive study of $f(R,T)$ gravity, the asymptotic implications of quadratic trace couplings in scalar field cosmology remain largely unexplored. We focus on a specific form given by $ f(R,T^\phi) = R + \alpha T^\phi + \beta (T^\phi)^2/\textcolor{black}{2}$, in which the parameters $\alpha$ and $\beta$ control the strength of non-minimal couplings between geometry and matter. We derive the set of cosmological equations for a spatially flat, homogeneous and isotropic universe and construct the autonomous system of first-order differential equations using a compact set of dimensionless variables. This formulation provides a foundation for the qualitative analysis of the asymptotic behavior. We identify and classify all critical points and analyze their stability properties. Finally, the energy conditions and the presence of dynamical instabilities are examined. We study the general scenario $\alpha \neq 0$ and $\beta \neq 0$, along with the subcases $\alpha = 0$ and $\beta = 0$, in order to compare with minimally coupled quintessence $\alpha = \beta = 0$. We find that the quadratic term in $T^\phi$ admits late-time accelerated de Sitter-like critical solutions at the background level. However, several accelerated points lie in a degenerate scalar sector with $Q_s=0$, where the standard linear perturbation criteria are inconclusive, while the quasi-de Sitter point with $Q_s>0$ is of saddle type. Therefore, establishing full perturbative viability requires going beyond the linear analysis in the degenerate sector.
	\end{abstract}
	
	
	
	\maketitle
	
	\section{Introduction}
	Observations of Type Ia supernovae (SNIa), Cosmic Microwave Background (CMB) anisotropies and Large-Scale Structure (LSS), to name a few, indicate that the universe is in a phase of accelerated expansion \cite{Perlmutter1998, Riess1998, Spergel2003, Adelman-McCarthy2006, Suzuki2012, Delubac2015, PlanckCollaboration2016}. To explain this late-time dynamical behavior within the framework of General Relativity (GR), theorists have proposed an exotic and negative pressure fluid known as Dark Energy (DE). The simplest candidate for DE is a positive cosmological constant $\Lambda$, with equation of state parameter $\omega_\Lambda=-1$, which accounts for approximately 70\% of the total energy content of the universe \cite{Carroll2001,Padmanabhan2003}. Besides, to account for local gravitational effects such as those observed in the rotation curves of galaxies and to provide a viable paradigm for structure formation, among other astrophysical observations, the introduction of a non-baryonic matter component, dubbed Dark Matter (DM) is still required \cite{Bertone2005,Capozziello2008}. This is known as the $\Lambda$CDM model and represents the  best-fitting picture for describing the majority of cosmological observations (see Refs. \cite{Peebles2003,Planck2020} for reviews ). Despite its success, the standard cosmological model remains theoretically unsatisfactory \cite{Zlatev1999, Ellis2014}. One of the most notable issues is the so-called \textit{cosmological constant problem}, which arises from the enormous discrepancy between the predicted and the observed value of the vacuum energy density \cite{Weinberg1989}. In addition, the fact that the energy densities of DE and DM are of the same order of magnitude at the present time requires extremely fine-tuned initial conditions in the early universe \cite{ Velten2014}. Finally, tensions persist between observations on both small and cosmological scales and this may indicate the existence of new physics \cite{Freedman2017,Verde2019,Riess2019}.

	On the other hand, the unknown nature of dark components has motivated alternative theories for cosmological dynamics. Some of these consist of either an alteration of the geometrical sector or by the addition of extra scalar, vector or tensor fields (see Refs. \cite{Copeland2006, Clifton2012,  Quiros2019r} for reviews). In the latter case, the simplest and most studied model is the canonical scalar field  minimally coupled to gravity $\phi$, which can be described by its kinetic term  $X \equiv - \nabla_\mu \phi \nabla^\mu \phi/2,$ and the self-interacting potential $V(\phi)$ \cite{Ratra1988, Wetterich1988, Caldwell1998}.  This family of models predicts an expanding accelerated universe that provides a nearly flat potential so that $X \ll V(\phi)$. Besides, they avoid -or at least alleviate- the fine tuning problem due to the existence of the so-called scaling solutions in which the scalar field kinetic and potential energies maintain a fixed ratio. The scalar field is characterized by a perfect fluid with equation of state $p_\phi = \omega_\phi \rho_\phi$, where $p_\phi$ denotes the pressure, $\rho_\phi$ corresponds to the energy density and $\omega_\phi$ is the equation of state parameter. For accelerated solutions, $\omega_\phi \leq -1/3$, is required. The field is dubbed \textit {quintessence} if $-1\leq \omega_\phi<-1/3$, $\omega_\phi = -1$ corresponds to the \textit{cosmological constant} scenario and the \textit{phantom} regime is characterized by condition $\omega_\phi<-1$.

	Alternatively, modified theories of gravity may provide a unified description of both early-time inflation and late-time cosmic acceleration without requiring the existence of both DE and DM. Among the most representative models are those in which the Einstein–Hilbert action is extended by promoting the scalar curvature $R$ to an arbitrary function, such as in $f(R)$ or $f(R,\mathcal{G})$ gravity, where $\mathcal{G}$ is the Gauss-Bonnet topological invariant, or even contributions from higher-order curvature terms \cite{Sharma2024,Chandra2025}. Another extension, namely $f(R,T)$ gravity, assumes an arbitrary function depending not only on the Ricci scalar but also on the trace of the stress-energy tensor $T$ \cite{Harko2011}. This family of models can be viewed as a generalization of $f(R, L_m)$ gravity, whereas the gravitational Lagrangian depends not only on the matter Lagrangian, but also on its variations with respect to the metric. The dependence on $T$ may be induced by exotic imperfect fluids arising from quantum effects, such as conformal anomalies, or may also be induced by relativistically covariant models of interacting DE. Considering that the covariant divergence of the stress-energy tensor generally does not vanish, the motion of massive particles becomes non-geodesic, occurring in the presence of an extra force orthogonal to the four-velocity \cite{Goncalves2022}. Astrophysical and cosmological applications of $f(R,T)$ gravity have been extensively explored in the literature \cite{Shamir2015,Singh2014,Jamil2012,Sharif2012,Deng2015,Sharif2013,Velten2017,Moraes2017}. 
	
	Inspired by the rich phenomenology of $f(R,T)$ gravity, we focus on the subclass $f(R,T^\phi)=R+F(T^\phi)$ and study the polynomial model $F(T^\phi)=\alpha T^\phi+\beta(T^\phi)^2/2$, where $\alpha$ and $\beta$ are constant parameters and $T^\phi$ denotes the trace of the scalar-field stress-energy tensor. We restrict to $\beta\ge 0$ and, unless otherwise stated, focus on the sector $\alpha\ge 0$. Observational constraints typically favor small trace couplings at late times, $|\alpha|\lesssim 10^{-3}$--$10^{-2}$, to preserve consistency with local gravity tests and the background expansion history \cite{Shabani2013,Moraes2016,Zaregonbadi2016}. Theoretical viability  requires the absence of pathological instabilities and consistency with early-universe constraints such as BBN as well as with the luminal propagation of tensor modes \cite{Uzan2011,Abbott2017}.
	
	This paper is organized as follows. In Section \ref{gravity}, we review the fundamentals of $f(R,T)$ gravity and derive the field equations for $f(R,T^\phi)= R + \alpha T^\phi + \beta (T^\phi)^2/2$. Assuming a FLRW background metric, in Section \ref{cosmology} we derive the system of cosmological equations and some relevant physical parameters. In Section \ref{dyn_sys}, the cosmological field equations are transformed into an equivalent dynamical system by introducing normalized variables. The critical points of the system are identified and their existence and stability conditions are determined. We compute the cosmologically relevant parameters, verify the conditions related to Laplacian and ghost instabilities, and evaluate the energy conditions. Section \ref{discussion} explores the cosmological implications of our results, while Section \ref{conclusions} summarizes our main findings. Throughout this work, we adopt natural units where $8 \pi G = c = 1$.

	\section{The $f(R,T^\phi)$ gravity} \label{gravity}
	The action of $f(R,T^\phi)$ gravity is depicted by
	\be
	S = \int d^{4}x\sqrt{-g} \left[\frac{1}{2} f(R,T^\phi) + \mathcal{L}_m + \mathcal{L}_\phi \right], \label{action}
	\ee
	where $g$ is the determinant of the metric tensor $g_{\mu \nu}$, $\mathcal{L}_m$ is the Lagrangian for the matter fields and $\mathcal{L}_\phi$ is the Lagrangian density for the scalar field. Here, $f(R,T^\phi)$ denotes an analytic function of both the Ricci scalar $R \equiv g^{\mu \nu} R_{\mu \nu}$ and the trace of the scalar field stress-energy tensor $T^\phi \equiv g^{\mu \nu} T_{\mu \nu}^\phi$. Varying action \eqref{action} with respect to the metric yields the field equations:
	\bea 
	f_R(R,T^\phi) R_{\mu \nu} - \frac{1}{2} g_{\mu \nu} f(R,T^\phi) + \left[g_{\mu \nu} \Box - \nabla_\mu \nabla_\nu \right] f_R(R,T^\phi) = T^m_{\mu \nu} + T^\phi_{\mu \nu} - f_T(R,T^\phi) \left(T^\phi_{\mu \nu} + \Theta^\phi_{\mu \nu}\right), \label{eq-of-mot}
	\eea
	with $f_R := \partial f(R,T^\phi)/\partial R$, $f_T := \partial f(R,T^\phi)/\partial T^\phi$, and $\Box \equiv g^{\mu \nu} \nabla_\mu \nabla_\nu$ is the d'Alembert operator. For the matter sector, the stress-energy tensor is defined as:
	\bea
	T_{\mu\nu}^m = - \frac{2}{\sqrt{-g}} \frac{\delta (\sqrt{-g} \mathcal{L}_m)}{\delta g^{\mu \nu}} = g_{\mu \nu} \mathcal{L}_m - 2 \frac{\delta \mathcal{L}_m}{\delta g^{\mu \nu}}, \label{tmunu}
	\eea
	and assuming the matter stress-energy tensor given by a perfect fluid:
	\be
	T_{\mu \nu}^m = (\rho_m + p_m) u_{\mu} u_{\nu} + p_m g_{\mu \nu},\ee
	where the four-velocity $u_\mu$ satisfies the condition $u^\mu u_\mu = -1$, and therefore $u^\nu \nabla_\mu u_\nu = 0$.
	
	In order to proceed, we must specify a particular form for $f(R,T^\phi)$.
	Motivated by the standard $f(R,T)$ framework, we focus on the  subclass
	\begin{equation}
		f(R,T^\phi)=R+F(T^\phi),
	\end{equation}
	where $F$ depends only on the trace of the scalar-field stress-energy tensor. Our restriction to the subclass $f(R,T^\phi)=R+F(T^\phi)$ follows a minimal-deformation strategy: the nonminimal trace interaction is confined to the scalar sector responsible for the accelerated dynamics, while the standard matter sector $\mathcal{L}_m$ remains minimally coupled to the metric. In particular, this avoids a direct trace coupling to ordinary matter (and the associated extra-force issues typical of $f(R,T)$ models with the total trace $T$), while still capturing departures from GR through the scalar trace $T^\phi$.
	
	Assuming that $F(T^\phi)$ is analytic, one may expand it around the GR limit as
	\begin{equation}
		F(T^\phi)=c_0+c_1T^\phi+c_2(T^\phi)^2+\mathcal{O}\!\left((T^\phi)^3\right).
	\end{equation}
	
	The constant term $c_0$ is degenerate (at the background level) with an effective cosmological constant and is therefore set to zero. Keeping the leading non-trivial terms, we adopt the minimal polynomial truncation
	\begin{equation}
		F(T^\phi)=\alpha T^\phi+\frac{\beta}{2}(T^\phi)^2,
	\end{equation}
	which captures the leading linear trace coupling and the first nonlinear correction. Similar polynomial trace couplings (including $R+\alpha T+\beta T^2$) have been widely employed to explore cosmological phenomenology and constrain parameters \cite{Baffou2019,Moraes2018}.
	
	In this work we take $\alpha\in\mathbb{R}$ and $\beta\ge 0$, and in the remainder we focus on the sector $\alpha\ge 0$, which includes the weak-coupling regime relevant for late-time cosmology and contains parameter regions where a non-degenerate, ghost-free branch can be obtained. Finally, the quadratic trace correction can be viewed as an effective nonlinear contribution associated with trace interactions, which may encode departures from GR induced by nonstandard matter effects  or quantum contributions such as conformal anomalies.
	
	For the canonical scalar field $\phi$ with a self-interacting potential $V(\phi)$, the Lagrangian is depicted by:
	\be
	\mathcal{L}_{\phi} = - \frac{1}{2} g^{\mu \nu} \nabla_\mu \phi \nabla_\nu \phi - V(\phi),
	\ee
	and its associated stress-energy tensor takes the form:
	\be
	T_{\mu \nu}^\phi = \nabla_\mu \phi \nabla_\nu \phi - g_{\mu \nu} \left[\frac{1}{2} (\nabla \phi)^2 + V(\phi) \right].
	\ee
	The trace is denoted by:
	\be
	T^\phi \equiv g^{\mu \nu} T_{\mu \nu}^\phi = - \Bigl[ (\nabla \phi)^2 + 4V(\phi) \Bigr],
	\ee
	and the variation of the stress-energy tensor  with respect to the metric takes the form:
	\be
	\Theta_{\mu \nu}^\phi = g^{\alpha \beta} \frac{\delta T_{\alpha \beta}^\phi}{\delta g^{\mu \nu}} = -2 T_{\mu \nu}^\phi + g_{\mu \nu} \mathcal{L}_\phi - 2 g^{\alpha \beta} \frac{\partial^2 \mathcal{L}_\phi}{\partial g^{\alpha \beta} \partial g^{\mu \nu}} = -2\,\nabla_{\mu}\phi\,\nabla_{\nu}\phi
	+ g_{\mu\nu}\left[\frac{1}{2}(\nabla\phi)^2+V(\phi)\right].
	\ee
	Finally, the Einstein field equation reads:
	\be
	G_{\mu\nu}
	= T^m_{\mu\nu} + \Big\{\,1+\alpha-\beta\big[(\nabla\phi)^2+4V(\phi)\big]\Big\}\nabla_\mu\phi\,\nabla_\nu\phi
	- g_{\mu\nu} \Bigg\{
	\frac{1}{2}(\nabla\phi)^2 + V(\phi)
	+ \frac{1}{2}\Big[
	\alpha\{(\nabla\phi)^2+4V(\phi)\}
	- \frac{\beta}{2}\{(\nabla\phi)^2+4V(\phi)\}^2
	\Big]\Bigg\}.
	\label{e10}
	\ee
	
	This choice reflects the focus on the dynamics of the scalar field as the driver of cosmic acceleration and emphasizes the interplay between the dynamics of the scalar field and the coupling constants $\alpha$ and $\beta$, which govern the deviations from standard gravity. Note that the GR limit is recovered for $\alpha=\beta=0$ (equivalently $F\to 0$), in which case Eq.~\eqref{eq-of-mot} reduces to the Einstein equations with a minimally coupled scalar field.

	
	\section{The cosmological Model} \label{cosmology}
	We assume a flat Friedmann-Robertson-Walker spacetime described by the line element 
	\be 
	ds^2 = -dt^2 + a^2(t) \delta_{ik} dx^i dx^k,
	\ee
	where $a(t)$ is the scale factor. The cosmological equations read:
	\bea
	3H^2 &=& \rho_m + \rho_\phi, \label{Fri_eq} \\
	-2\dot{H} &=& \rho_m + \rho_\phi + p_\phi, \label{Ray_eq} \\
	\dot{\rho}_m + 3H \rho_m &=& 0, \label{mat_eq} \\
	(1 + \alpha - 4\beta V + 3\beta \dot{\phi}^2) \ddot{\phi} + 3(1 + \alpha - 4\beta V) H \dot{\phi} &+& 3\beta H \dot{\phi}^3 = (2\beta \dot{\phi}^2 + 8\beta V - 2\alpha - 1) V_\phi, \label{KG_eq}
	\eea
	where $H := \dot{a}/a$ is the Hubble parameter, $V_\phi = \partial V / \partial \phi$ and dots denote derivatives with respect to $t$. In addition to the scalar field $\phi$, we assume the presence of a pressureless matter fluid characterized by an energy density $\rho_m$. The energy density and pressure for the scalar field take the form:
	\begin{align}
		\rho_\phi &= \frac{3}{4} \beta \dot{\phi}^4 + \frac{ \dot{\phi}^2}{2}\left(1 + \alpha - 4\beta V \right) + (1 + 2\alpha - 4\beta V)V, \\
		p_\phi &= \frac{1}{4} \beta \dot{\phi}^4 + \frac{ \dot{\phi}^2}{2}(1 + \alpha - 4\beta V) - (1 + 2\alpha - 4\beta V)V, \label{randp}
	\end{align}
	which are related through the DE equation of state:
	\begin{equation}
		\omega_\phi = \frac{p_\phi}{\rho_\phi}.
	\end{equation}
	In what follows, $\rho_\phi$ and $p_\phi$ denote the \emph{effective} energy density and pressure associated with the scalar sector in the $f(R,T^\phi)$ field equations, given by Eqs. \eqref{randp}. They reduce to the standard minimally coupled expressions when $\alpha=\beta=0$.
	Here we assume a specific form for the self-interacting potential of the scalar field. One of the simplest and well-motivated choices is the quadratic form:
	\begin{equation}
		V(\phi) = \frac{1}{2}m^2 \phi^2,
	\end{equation}
	where $m$ denotes the mass of the scalar field. This potential naturally emerges as the lowest-order expansion of a general potential around $\phi = 0$ \cite{Starobinskii1978,Belinsky1985} and is well motivated by both particle physics and by inflationary cosmology, where massive scalar fields play a key role in the early-universe dynamics \cite{Sahni2000,Martin2024}. The quadratic potential provides a smooth transition from a matter-dominated to a DE–dominated phase, consistent with CMB and LSS observations. Dynamical system analyses have shown that this potential can yield cosmologically viable trajectories, provided that $m \sim 10^{-33}$ $\text{eV}$, ensuring field evolution on cosmological timescales \cite{Linder2006,Marsh2015}. For a quadratic potential the dynamics can be written as a three-dimensional autonomous system \cite{Urena2009} and for non-minimal coupling frameworks the quadratic form still plays a prominent role \cite{Hrycyna2007}.
	
	It is important to notice that exponential potential, $V(\phi)=V_0 e^{-\lambda\phi}$ with constant $\lambda$, is frequently employed in late-time quintessence models and inflationary scenarios, since it reduces the dimensionality of the autonomous system by fixing $V_\phi/V=-\lambda$ and produces scaling solutions \cite{Amendola2000}. In contrast, we adopt the quadratic form in order to gain an additional dynamical degree of freedom and thus capture a richer phase-space structure. This choice, typical of early-time inflation and massive scalar field cosmologies, also enables the study of transitions between decelerated and accelerated regimes within the same framework.
	
	Finally, since any viable cosmological model requires both decelerated and accelerated phases of expansion to describe the entire evolutionary history of the universe, it is necessary to introduce a dimensionless measure of cosmic acceleration. The deceleration parameter
	\begin{equation}
		q := -1 - \frac{\dot{H}}{H^2},
	\end{equation}
	indicates that the expansion accelerates if $q < 0$, although $q > 0$ corresponds to a deceleration. 
	
	
	\subsection{Energy conditions}
	
	Energy conditions play a major role in gravitational physics, as they provide model-independent criteria for the physical viability of matter fields \cite{Rubakov2014}. In the context of GR, they are essential for deriving singularity theorems \cite{Hawking1970}, analyzing the causal structure of spacetime and investigating the attractive nature of gravity through the Raychaudhuri equation \cite{Poisson2004}. When scalar fields are present and non-minimally coupled to curvature or matter terms, these conditions are significantly modified. In this scenario, the presence for both linear and quadratic couplings to the trace of the stress-energy tensor alters the expressions for energy density and pressure, leading to modified conditions. The classical energy conditions are defined as follows:
	\begin{itemize}
		\item Null Energy Condition (NEC): $T_{\mu\nu} k^\mu k^\nu \geq 0$ for all null vectors $k^\mu$, reduces to $\rho_\phi + p_\phi \geq 0, \,\,\Rightarrow \,\,  (1 + \alpha - 4\beta V + \beta \dot{\phi}^2)\dot{\phi}^2 \geq 0$.
		\item Weak Energy Condition (WEC): $T_{\mu\nu} u^\mu u^\nu \geq 0$ for all time-like vectors $u^\mu$, requires $\rho_\phi \geq 0$ and $\rho_\phi + p_\phi \geq 0$,which in our model read: $\frac{3}{4} \beta \dot{\phi}^4 + \frac{ \dot{\phi}^2}{2}\left(1 + \alpha - 4\beta V \right) + (1 + 2\alpha - 4\beta V)V\geq 0$ and $(1 + \alpha - 4\beta V + \beta \dot{\phi}^2)\dot{\phi}^2 \geq 0$.
		\item  Dominant Energy Condition (DEC): $\rho_\phi \geq |p_\phi|,$ which translates to $\rho_\phi + p_\phi = (1 + \alpha - 4\beta V + \beta \dot{\phi}^2)\dot{\phi}^2 \geq 0 \quad$ and $\rho_\phi - p_\phi  = \beta\dot{\phi}^4/2 + 2(1 + 2\alpha - 4\beta V)V \geq 0 $.
		\item Strong Energy Condition (SEC): $\rho_\phi + 3 p_\phi \geq 0,$ then  $3\beta \dot{\phi}^4/2 + 2(1 + \alpha - 4\beta V)\dot{\phi}^2 - 2(1 + 2\alpha - 4\beta V)V \geq 0$.
	\end{itemize}

	Violation of the SEC is naturally associated with late-time cosmic acceleration and typically occurs at de Sitter critical points. On the other hand, NEC and WEC are generally satisfied in weakly coupled regimes, but may be violated in specific regions of phase space when $\beta$ is large ($z \ll 1$). Finally, the DEC condition provides bounds on the effective pressure and its violation may indicate the presence of stiff or exotic DE components. 
	
	In this analysis, the energy conditions are evaluated for the effective scalar field only, since the matter sector is minimally coupled and separately conserved $\dot{\rho}_m + 3H\rho_m = 0$. If one wishes to apply the conditions to the total fluid, one simply replaces $\rho_{\rm tot} = \rho_m + \rho_\phi$ and $p_{\rm tot} = p_\phi$ (with $p_m = 0$). This addition does not affect the qualitative behavior, since NEC, WEC and DEC remain satisfied in the same regions, while SEC violation persists in the scalar-field–dominated regimes.
	

	\subsection{On ghosts, Laplacian stability and causality issues}
	In order to discuss the stability of theories described by the action \eqref{action} in the cosmological context, it is necessary to consider linear perturbations around a flat FLRW background. This approach provides an opportunity to investigate their effects on cosmic growth as well on the modifications introduced to the generalized Poisson equation. The equations of motion for the field perturbations take the form of a generalized wave equation. To ensure physical viability, the following conditions must be satisfied \cite{DeFelice2012} \footnote{At the background and linear level, the model $f(R,T^\phi)=R+\alpha T^\phi+\beta (T^\phi)^2/2$ can be recast as an effective k-essence theory of the Horndeski type, with $G_4=1/2$, $G_3=G_5=0$, and $G_2=K(\phi,X)=(1+\alpha)X-(1+2\alpha)V+\beta(X^2-4XV+4V^2)$.}
	
	\begin{itemize}
		\item Absence of ghosts: The kinetic coefficient of scalar perturbations must be strictly positive in order to avoid ghost modes (negative kinetic terms in the Hamiltonian). This requires:
		\begin{equation}
			Q_s = \frac{w_1 (4w_1 w_3 + 9w_2^2)}{3w_2^2} > 0. \label{ghost}
		\end{equation}
		The limiting case $Q_s=0$ corresponds to a degenerate (vanishing) kinetic term for scalar perturbations, which signals a marginal/strong-coupling regime where the standard linear perturbation analysis becomes inconclusive. Therefore, fixed points with $Q_s=0$ are classified as marginal from the perturbative perspective.  Throughout the paper, we only label a fixed point as ghost-free when $Q_s>0$. For $Q_s=0$, the scalar sector is degenerate and the no-ghost condition is inconclusive at the level of linear perturbations; we therefore report such cases as marginal/strong-coupling \cite{DeFelice2010, DeFelice2012b, Leon2013}.
		
		\item Absence of Laplacian instabilities: The speed of propagation of the perturbations $c_s^2 $, must satisfy:
		\begin{equation}
			c_s^2 = \frac{3(2w_1^2 w_2 H - w_2^2 w_4 - 2w_1^2 \dot{w}_2) - 6w_1^2 \rho_m}{w_1(4w_1 w_3 + 9w_2^2)}, \quad 0 \leq c_s^2 \leq 1, \label{cs}
		\end{equation}
		where $ c_s^2 \geq 0 $ ensures the absence of Laplacian instabilities and $ c_s^2 \leq 1 $ guarantees causality.
	\end{itemize}
	
	The coefficients $w_1, w_2, w_3$ and $w_4$ quantify contributions from the kinetic and potential terms of the scalar field, as well as corrections from the coupling constants $\alpha$ and $\beta$. For the theory described by the action \eqref{action}, the coefficients $ w_i $ read:
	$$ w_1 = w_4 = 1,  \quad  w_2 = 2H, \quad \dot{w}_2 = 2 \dot H, \quad w_3 = 3 X (K_X + 2X K_{XX}) -9H^2,$$
	where $K_X = \partial K/\partial X$. For tensor perturbations, the conditions for avoiding ghosts and Laplacian instabilities in this case are depicted by:
	\begin{equation}
		Q_T \equiv \frac{w_1}{4} \geq 0, \quad c_T^2 \equiv \frac{w_1}{w_4} \geq 0, \label{Qs}
	\end{equation}
	and it is straightforward to verify that both conditions are trivially satisfied for our model. Finally, vector perturbations decay rapidly in an expanding universe, making their contribution dynamically negligible for large-scale cosmological evolution \cite{Linder2012}.

	
	\section{The dynamical system} \label{dyn_sys}
	Our aim in this Section is to trade the system of second-order equations \eqref{Fri_eq}--\eqref{randp} into an autonomous system of first-order differential equations by introducing a bounded set of dimensionless variables. This enables the study of the global asymptotic properties of the model by means of dynamical systems techniques. The resulting system takes the form
	$$ \dot{\vec{x}} = \vec{f}(\vec{x}),$$
	where $\vec{x}$ is the state vector in phase space, the overdot denotes differentiation with respect to cosmic time and $\vec{f}(\vec{x})$ encodes the dynamics of the system. Critical points $\vec{x}_c$ are defined by $\vec{f}(\vec{x}_c) = \vec{0}$, and their stability is determined by studying small perturbations around equilibrium, i.e., $\vec{x} = \vec{x}_c + \vec{u}$. According to the Hartman-Grobman theorem \cite{Hartman1960,Grobman1962}, the dynamics near a hyperbolic fixed point (i.e. $\Re(\lambda_i)\neq 0$ for all eigenvalues of the Jacobian $\mathbb{M}$) is topologically equivalent to its linearization
	\be
	\dot{\vec{u}} = \mathbb{M} \vec{u}.
	\ee
	Hence, for hyperbolic points the sign of $\Re(\lambda_i)$ determines local stability. For \emph{non-hyperbolic} fixed points (at least one eigenvalue with $\Re(\lambda)=0$), linearization is inconclusive and one must resort to a center-manifold analysis or complementary numerical integration. In what follows, such cases are explicitly labeled as non-hyperbolic and their local behavior is assessed numerically.
	
	Following \cite{Dearcia2016}, we define the dimensionless variables:
	\begin{equation}
		x_\pm := \frac{\sqrt{6}H}{\dot{\phi} \pm \sqrt{6}H}, \quad
		y := \frac{\sqrt{6}H}{m \phi + \sqrt{6}H}, \quad
		z := \frac{1}{H^2 \beta + 1}, \quad
		u := \frac{H}{m + H}, \label{xyz_var}
	\end{equation}
	where, $x_+$ corresponds to $\dot{\phi} \geq 0$ and $x_-$ to $\dot{\phi} \leq 0$, assuming $\dot{\phi}$ does not change sign along phase space orbits. We also restrict to expanding cosmologies ($H \geq 0$). The limit $z \to 1$ corresponds to the regime where the coupling $\beta$ becomes negligible ($\beta \ll H^{-2}$), while $z = 0$ defines the boundary surface where coupling effects dominate. Finally, $u=0$ does not necessarily imply $H=0$, but it is also true for $H \ll m$. Similarly, $u=1$ holds both for $H \rightarrow \infty$ and $H \gg m$.
	
	From \eqref{Fri_eq}, the matter density is eliminated as
	\be
	\rho_m = 3H^2 - \rho_\phi,
	\label{rm_elim}
	\ee
	then, the matter fraction reads
	\begin{equation}
		\Omega_m := \frac{\rho_m}{3H^2} = 1 - \frac{\rho_\phi}{3H^2}.
		\label{Om_def}
	\end{equation}
	In all background relations used to build the autonomous system, $\rho_m$ is replaced by \eqref{rm_elim}, so no independent matter variable is required. In terms of the set of variables \eqref{xyz_var}, the cosmological equations transform into the following autonomous system:
	\begin{align}
		x'_\pm &= x_\pm (1 \mp x_\pm) \left( \frac{\dot{H}}{H^2} \right)_\pm - \frac{x_\pm^2}{\sqrt{6}} \left( \frac{\ddot{\phi}}{H^2} \right)_\pm, \label{ode1} \\
		y' &= y (1 - y) \left( \frac{\dot{H}}{H^2} \right)_\pm - y^2 \left( \frac{1 \mp x_\pm}{x_\pm} \right) \left( \frac{1 - u}{u} \right), \label{ode2} \\
		z' &= -2 z (1 - z) \left( \frac{\dot{H}}{H^2} \right)_\pm, \label{ode3} \\
		u' &= u (1 - u) \left( \frac{\dot{H}}{H^2} \right)_\pm, \label{ode4}
	\end{align}
	where prime denotes derivative with respect to $N = \ln a$, and
	\begin{align}
		\left( \frac{\dot{H}}{H^2} \right)_\pm = - \frac{3}{2} \Bigg\{ 
		1 + (1 + \alpha) \left( \frac{1 \mp x_\pm}{x_\pm} \right)^2 
		- (1 + 2\alpha) \left( \frac{1 - y}{y} \right)^2 \nonumber \\
		+ 3 \left( \frac{1 - z}{z} \right) \Bigg[
		\left( \frac{1 \mp x_\pm}{x_\pm} \right)^4 
		- 4 \left( \frac{1 \mp x_\pm}{x_\pm} \right)^2 \left( \frac{1 - y}{y} \right)^2 
		+ 4 \left( \frac{1 - y}{y} \right)^4 
		\Bigg] \Bigg\}, \label{Hdot}
	\end{align}
	and
	\begin{align}
		\left( \frac{\ddot{\phi}}{H^2} \right)_\pm = 
		&- \Bigg\{ \sqrt{6} \left( \frac{1 - y}{y} \right) \left( \frac{1 - u}{u} \right) 
		\left[ 1 + 2\alpha - 24 \left( \frac{1 - y}{y} \right)^2 \left( \frac{1 - z}{z} \right) 
		+ 12 \left( \frac{1 \mp x_\pm}{x_\pm} \right)^2 \left( \frac{1 - z}{z} \right) \right] \nonumber \\
		&+ 3\sqrt{6} \left( \frac{1 \mp x_\pm}{x_\pm} \right) 
		\left[ 1 + \alpha - 12 \left( \frac{1 - y}{y} \right)^2 \left( \frac{1 - z}{z} \right) 
		+ 6 \left( \frac{1 \mp x_\pm}{x_\pm} \right)^2 \left( \frac{1 - z}{z} \right) \right] \nonumber \\
		& - 24\sqrt{6} \left(\frac{1-u}{u}\right)
		\left(\frac{1 \mp x_\pm}{x_\pm}\right)^2
		\left(\frac{1-y}{y}\right) \left(\frac{1-z}{z}\right) \Bigg\} \Bigg/ \nonumber \\
		&\Bigg[ 1 + \alpha + 18 \left( \frac{1 \mp x_\pm}{x_\pm} \right)^2 \left( \frac{1 - z}{z} \right) 
		- 12 \left( \frac{1 - y}{y} \right)^2 \left( \frac{1 - z}{z} \right) \Bigg]. \label{phiddot}
	\end{align}
	The modified Friedmann constraint becomes:
	\begin{align}
		\Omega_m = 1 
		&- (1 + \alpha) \left( \frac{1 \mp x_\pm}{x_\pm} \right)^2 
		- (1 + 2\alpha) \left( \frac{1 - y}{y} \right)^2 \nonumber \\
		&- 3 \left( \frac{1 - z}{z} \right) \left[ 3 \left( \frac{1 \mp x_\pm}{x_\pm} \right)^4 
		- 4 \left( \frac{1 - y}{y} \right)^2 \left( \frac{1 \mp x_\pm}{x_\pm} \right)^2 
		- 4 \left( \frac{1 - y}{y} \right)^4 \right],
	\end{align}
	and the physically meaningful region of the phase space is defined by:
	\begin{equation}
		\Psi = \Psi^+ \cup \Psi^-, \quad \Psi^\pm = \left\{ (x_\pm, y, z, u) \, \bigg| \, |x_\pm| \leq 1,\, 0 \leq y \leq 1,\, 0 \leq z \leq z_{surf},\, 0 \leq u \leq 1 \right\}, \label{ps}
	\end{equation}
	where
	\begin{equation}
		z_{surf} = \frac{3 \left[ 3 \left( \frac{1 \mp x_\pm}{x_\pm} \right)^4 - 4 \left( \frac{1 - y}{y} \right)^2 \left( \frac{1 \mp x_\pm}{x_\pm} \right)^2 - 4 \left( \frac{1 - y}{y} \right)^4 \right]}{3 \left[ 3 \left( \frac{1 \mp x_\pm}{x_\pm} \right)^4 - 4 \left( \frac{1 - y}{y} \right)^2 \left( \frac{1 \mp x_\pm}{x_\pm} \right)^2 - 4 \left( \frac{1 - y}{y} \right)^4 \right] + 1 - (1 + \alpha) \left( \frac{1 \mp x_\pm}{x_\pm} \right)^2 - (1 + 2\alpha) \left( \frac{1 - y}{y} \right)^2}.
	\end{equation}
	
	\subsection{General scenario: $\alpha \neq 0$, $\beta \neq 0$} \label{general}
	A general scenario within the framework of $ f(R,T^\phi) = R + \alpha T^\phi + \beta (T^\phi)^2/2$ gravity, includes both linear and quadratic couplings between $T^\phi$ and the geometry. We analyze the behavior of the matter and DE sectors through the density parameter $\Omega_m$, the scalar field equation of state $\omega_\phi$ and the deceleration parameter $q$. In addition, we also evaluate the conditions for ghost and Laplacian instabilities $Q_s$ and $c_s^2$, along with the energy conditions.
	
	\begin{itemize}
		\item Points $p_1^\pm = (\pm 1, 1, 0, 1)$: These correspond to matter-dominated big-bang-type saddle points. The matter density parameter satisfies $\Omega_m = 1$, and the scalar field equation of state $\omega_\phi$ remains undetermined. The deceleration parameter takes the value $q = \tfrac{1}{2}$, corresponding to a decelerated expansion. The eigenvalues of the linearized system are depicted by $\{-3, -\frac{3}{2}, \tfrac{3}{2}, und\}$, indicating a saddle point. The solution is free of Laplacian instabilities $c_s^2=1$, while the scalar sector is degenerate $Q_s=0$; therefore the no-ghost condition is inconclusive at linear order. All the energy conditions are saturated $\rho_\phi=p_\phi=0$. 
		\item Points $p_2^\pm = (\pm 1, 1, 1, 1)$: These are matter-dominated saddle points with $z = 1$, which indicates that the nonlinear coupling $\beta$ becomes negligible $\beta \ll H^{-2}$, while $u = 1$ corresponds to $m \ll H$. The matter density entirely dominates with $\Omega_m = 1$ and undetermined equation of state $\omega_\phi$. The deceleration parameter is $q = \tfrac{1}{2}$, which is consistent with a standard matter-dominated era. The eigenvalues of the linearized system are $\{-3, -\tfrac{3}{2}, \tfrac{3}{2}, \tfrac{3}{2}\}$, indicating a saddle point. The solution is free of Laplacian instabilities $c_s^2=1$, while the scalar sector is degenerate  $Q_s=0$. As in the previous case, all energy conditions are saturated.
		\item Points $p_3^\pm = (\pm 1, \tfrac{1}{2}, \tfrac{6}{6 + \alpha}, 1)$: These correspond to scalar field-dominated de Sitter attractors. The universe undergoes a exponential acceleration phase with $q = -1$, $\omega_\phi = -1$, while the matter sector is suppressed $\Omega_m = 0$. A numerical analysis shows that these are non-hyperbolic stable attractors with eigenvalues $\{-3,-3,0,0\}$. The solution is free of Laplacian instabilities $c_s^2=1$, while the scalar sector is degenerate  $Q_s=0$. Here, NEC and DEC are saturated, WEC is satisfied and SEC is violated.
		
		\item Points $p_4^\pm = \left(\pm 1, \frac{\sqrt{1+2\alpha}}{\sqrt{1+2\alpha}+\sqrt{2}}, \frac{48}{49+4\alpha+4\alpha^2}, u\right)$: These correspond to a family of scalar field-dominated de Sitter solutions, with $q = -1$, $\omega_\phi = -1$ and $\Omega_m = 0$. The critical point is a non-hyperbolic saddle, with eigenvalues $\{0,-3,f_1(\alpha,u),f_2(\alpha,u)\}$, with the specific property that for every $\alpha>0$, $f_1(\alpha,u)$ is always nonnegative and $f_2(\alpha,u)$ is always nonpositive. The solution is free of Laplacian instabilities $c_s^2=1$, while the scalar sector is degenerate  $Q_s=0$. Just like in the previous scenario, NEC and DEC are saturated, while WEC is satisfied and SEC is violated.
		\item Points $p_5^\pm = \left(\pm \frac{1}{2}, \frac{2\alpha}{2\alpha + \sqrt{2\alpha(3+\alpha)}}, \frac{18}{18+\alpha+\alpha^2}, 1 \right)$: These critical points exist only for $\alpha \neq 0$. They correspond to matter vacuum scalar field-dominated solutions $\Omega_m = 0$ and exhibit accelerated expansion with $\omega_\phi = -1$ and $q = -1$, respectively. The eigenvalues of the linearized system reads $\{ 0,-3,-3,0\}$, and therefore linear stability is inconclusive. A numerical inspection of nearby trajectories suggests saddle-like behavior. As in solutions $p_3^\pm$ and $p_4^\pm$, NEC and DEC are saturated, WEC is satisfied and SEC is violated. The solution is free of ghost and Laplacian instabilities, since $Q_s=2 \alpha (1 + \alpha)$ and  $c_s^2=0$, respectively.
		\item Points $p_6^\pm = \left(\frac{1 + \alpha \mp \sqrt{1 + \alpha}}{\alpha}, 1, 1, 1 \right)$: These points exist only for $\alpha \neq 0$ and correspond to decelerated stiff fluid solutions, with $\omega_\phi = 1$ and $q = 2$. The universe is entirely dominated by the scalar field $\Omega_m = 0$ and behaves as saddles with eigenvalues $\{-6,3,3,3\}$ . Ghost and Laplacian instabilities are absent, since $Q_s = 3$ and $c_s^2 = 1$. Here, NEC, WEC and SEC are satisfied, while DEC is saturated.
		
	\end{itemize}
	
	\begin{table*}[htbp]
		\centering
		\begin{tabular}{||c||c|c|c|c|c|c|c|c|c|c|c|c||}
			\hline\hline
			Crit. Point & $x_\pm$ & $y$ & $z$ & $u$ & Existence & $\Omega_m$ & $q$ & $\omega_\phi$ & NEC & WEC & DEC & SEC \\
			\hline\hline
			$p_1^\pm$ & $\pm 1$ & $1$ & $0$ & $1$ & always & $1$ & $1/2$ & und. & saturated & saturated & saturated & saturated \\
			$p_2^\pm$ & $\pm 1$ & $1$ & $1$ & $1$ & always & $1$ & $1/2$ & und. & saturated & saturated & saturated & saturated  \\
			$p_3^\pm$ & $\pm 1$ & $1/2$ & $\frac{6}{6+\alpha}$ & $1$ & always & $0$ & $-1$ & $-1$ &  saturated & satisfied & saturated  & violated \\
			$p_4^\pm$ & $\pm 1$ & $\frac{\sqrt{1+2\alpha}}{\sqrt{1+2\alpha}+\sqrt{2}}$ & $\frac{48}{49+4\alpha+4\alpha^2}$ & $u$ & always & $0$ & $-1$ & $-1$ & saturated & satisfied &saturated & violated  \\
			$p_5^\pm$ & $\pm 1/2$ & $\frac{2\alpha}{2\alpha + \sqrt{2\alpha(3+\alpha)}}$ & $\frac{18}{18+\alpha+\alpha^2}$ & $1$ & $\alpha \neq 0$ & $0$ & $-1$ & $-1$ & saturated & satisfied & saturated & violated \\
			$p_6^\pm$ & $\frac{1+\alpha \mp \sqrt{1+\alpha}}{\alpha}$ & $1$ & $1$ & $1$ & $\alpha \neq 0$ & $0$ & $2$ & $1$ & satisfied & satisfied & saturated & satisfied \\
			\hline\hline 
		\end{tabular}
		\caption{Physically meaningful critical points of the autonomous system \eqref{ode1}--\eqref{ode4}, along with their existence conditions and cosmological parameters. The energy conditions (NEC, WEC, DEC, SEC) are also evaluated. Here, ``und.'' denotes a 0/0 indeterminacy and a condition is said to be \textit{satisfied} when the corresponding inequality holds, \textit{violated} when it is reversed and \textit{saturated} when equality is met (e.g. $\rho_\phi+p_\phi=0$ for NEC or $\rho_\phi=|p_\phi|$ for DEC).}
		\label{tab-1}
	\end{table*}
	
	\begin{table*}[htbp]
		\centering
		\begin{tabular}{||c||c|c|c|c|c|c|c||}
			\hline\hline
			Crit. Point & $\lambda_1$ & $\lambda_2$ & $\lambda_3$ & $\lambda_4$ & Stability & $Q_s$ & $c_s^2$ \\
			\hline\hline
			$p_1^\pm$ & $-3$ & $-3/2$ & $3/2$ & und. & saddle & $0$ & $1$ \\
			$p_2^\pm$ & $-3$ & $-3/2$ & $3/2$ & $3/2$ & saddle & $0$ & $1$ \\
			$p_3^\pm$ & $-3$ & $-3$ & $0$ & $0$ & non-hyperbolic attractor (num. inv.) & $0$ & $1$ \\
			$p_4^\pm$ & $0$ & $-3$ & $f_1(\alpha,u)$ & $f_2(\alpha,u)$ & saddle (num. inv.) & $0$ & $1$ \\
			$p_5^\pm$ & $0$ & $-3$ & $-3$ & $0$ & saddle (num. inv.) & $2\alpha(1+\alpha)$ & $0$ \\
			$p_6^\pm$ & $-6$ & $3$ & $3$ & $3$ & saddle & $3$ & $1$ \\
			\hline\hline
		\end{tabular}
		\caption{Eigenvalues of the linearized system of the autonomous system \eqref{ode1}--\eqref{ode4} at each critical point and corresponding stability classification. Conditions for ghost ($Q_s$) and Laplacian ($c_s^2$) stability are also shown. Here, $f_1(\alpha,u)$ and $f_2(\alpha,u)$ are high-complex functions.}
		\label{tab-2}
	\end{table*}
	
	These results emphasize the rich phase-space structure of $f(R,T^\phi)$ models with linear and quadratic trace couplings. The system admits matter-dominated saddles, stiff-fluid regimes and accelerated (de Sitter-like) scalar-field-dominated solutions. At the perturbative level, several accelerated fixed points satisfy $c_s^2\ge 0$ but lie in a degenerate scalar sector with $Q_s=0$, where the standard linear no-ghost test becomes inconclusive. In contrast, the quasi-de Sitter point $p_5^\pm$ can be genuinely ghost-free for $\alpha>0$ ($Q_s>0$) while featuring $c_s^2=0$. Therefore, while the model can realize late-time acceleration at the background level, establishing full perturbative viability and confronting the parameter space with observations requires going beyond the linear analysis, particularly in the degenerate sector.
	
	\subsection{Case $\alpha=0$}
	
	In the absence of linear coupling ($\alpha = 0$), the theory reduces to $f(R,T^\phi) = R + \frac{1}{2} \beta (T^\phi)^2$, isolating the effects of the purely quadratic trace interaction. This scenario provides an ideal test environment to probe the impact of higher-order matter-curvature couplings on the cosmological dynamics.
	
	\begin{itemize}
		\item Points $q_1^\pm = (\pm 1, 1, 0, 1)$:
		These are matter-dominated big-bang-type saddle points with $\Omega_m = 1$, while the scalar field equation of state $\omega_\phi$ remains undetermined. The expansion is decelerated with $q = \tfrac{1}{2}$. The eigenvalues of the linearized system are $\{-3, -\tfrac{3}{2}, \tfrac{3}{2},und.\}$, indicating a non-hyperbolic saddle. The solution is free of Laplacian instabilities ($c_s^2=1$), while the scalar sector is degenerate  ($Q_s=0$). All the energy conditions are saturated $\rho_\phi = p_\phi = 0$. These points correspond to early universe configurations, analogous to the standard big-bang matter-dominated phase and they are always present regardless of the value of $\beta$.
		
		\item Points $q_2^\pm = (\pm 1, 1, 1, 1)$:
		These saddle points share all physical properties with $q_1^\pm$, but correspond to a different region of the phase space $z=1$, where the quadratic trace coupling becomes negligible. We have a decelerated matter dominated solution $\Omega_m = 1$, $q = 1/2$, and undefined $\omega_\phi$. The eigenvalues are depicted by $\{-3, -\tfrac{3}{2}, \tfrac{3}{2}, \tfrac{3}{2}\}$, indicating a saddle character. The solution is free of Laplacian instabilities ($c_s^2=1$), while the scalar sector is degenerate  ($Q_s=0$). All the energy conditions are saturated.
		
		\item Points $q_3^\pm = (\pm 1, \tfrac{1}{2}, 1, 1)$:
		These correspond to scalar field-dominated de Sitter attractors, characterized by $\Omega_m = 0$, $\omega_\phi = -1$ and a fully accelerated expansion with $q = -1$. The eigenvalues are $\{-3, -3, 0, 0\}$, indicating a non-hyperbolic attractor (numerical investigation). The solution is free of Laplacian instabilities ($c_s^2=1$), while the scalar sector is degenerate ($Q_s=0$). Here, NEC and DEC are saturated, WEC is satisfied and SEC is violated. 
		
		\item Points $q_4^\pm = (\pm \tfrac{1}{2}, 1, 1, 1)$:
		Stiff-fluid saddle points dominated by the scalar field $\Omega_m = 0$, with equation of state $\omega_\phi = 1$ and a decelerated expansion $q = 2$. The eigenvalues are $\{-6, 3, 3, 3\}$, confirming a saddle and the ghost and Laplacian instabilities are absent $Q_s = 3$ and $c_s^2 = 1$. The energy conditions NEC and WEC are satisfied, while DEC is saturated and SEC is satisfied. These solutions represent kinetic-energy-dominated cosmologies, which are generally transient and do not provide late-time acceleration.
		
		\item Points $q_5^\pm = (\pm 1, \sqrt{2}-1, \tfrac{48}{49}, u)$:
		This is a one-parameter family of scalar field-dominated de Sitter-type saddle points. The expansion is accelerated $q = -1$ and the field behaves as a cosmological constant $\omega_\phi = -1$ and $\Omega_m = 0$. The eigenvalues are $\{-3, -3, 0, 0\}$, indicating a non-hyperbolic saddle character (numerical investigation). However, by applying numerical methods it can be found their saddle nature. The solution is free of Laplacian instabilities ($c_s^2=1$), while the scalar sector is degenerate  ($Q_s=0$). Here, NEC and DEC are saturated, WEC is satisfied while the SEC is violated. This family generalizes the late-time attractors, revealing the richness of the phase space even in the purely quadratic case. SEC violation again supports the emergence of cosmic acceleration.
	\end{itemize}
	
	\begin{table*}\centering
		\begin{tabular}{||c||c|c|c|c|c|c|c|c|c|c|c|c||}
			\hline\hline
			Crit. Point  & $x_\pm$ & $y$ & $z$ & $u$ & Existence & $\Omega_m$ & $q$ & $\omega_\phi$ & NEC & WEC & DEC & SEC \\
			\hline\hline
			$q_1^\pm$ & $\pm 1$ & $1$ & $0$ & $1$ & always & 1 & 1/2 & und. & saturated & saturated & saturated & saturated \\
			$q_2^\pm$ & $\pm 1$ & $1$ & $1$ & $1$ & always & $1$ & $1/2$ & und. & saturated & saturated &saturated & saturated \\
			$q_3^\pm$ & $\pm 1$ & $\frac{1}{2}$ & $1$ & $1$ & always & $0$ & $-1$ & $-1$ & saturated & satisfied &saturated & violated \\
			$q_4^\pm$ & $\pm \frac{1}{2}$ & $1$ & $1$ & $1$ & always & $0$ & $2$ & $1$ & satisfied & satisfied & saturated & satisfied \\
			$q_5^\pm$ & $\pm 1$ & $\sqrt{2}-1$ & $\frac{48}{49}$ & $u$ & always & $0$ & $-1$ & $-1$ & saturated & satisfied & saturated & violated \\
			\hline\hline
		\end{tabular}
		\caption{Special case $\alpha=0$. Physically meaningful critical points of the autonomous system \eqref{ode1}--\eqref{ode4}, with their existence conditions and relevant cosmological parameters, including energy conditions (NEC, WEC, DEC, SEC).}
		\label{tab-3}
	\end{table*}
	
	\begin{table*}\centering
		\begin{tabular}{||c||c|c|c|c|c|c|c||}
			\hline\hline
			Crit. Point & $\lambda_1$ & $\lambda_2$ & $\lambda_3$ & $\lambda_4$ & Stability  & $Q_s$& $c_s^2$\\
			\hline\hline
			$q_1^\pm$ & $-3$ & $-3/2$ & 3/2 & und. &  saddle & $0$ & $1$\\
			$q_2^\pm$  & $-3$ & $-3/2$ & $3/2$ & $3/2$ &  saddle & $0$ & $1$\\
			$q_3^\pm$  & $-3$ & $-3$ & $0$ & $0$ &  non-hyperbolic attractor (num. inv.)& $0$ & $1$\\
			$q_4^\pm$  & $-6$ & $3$ & $3$ & $3$ &  saddle & $3$ & $1$\\
			$q_5^\pm$ & $-3$ & $-3$ & $0$ & $0$ &  saddle & $0$ & $1$\\
			\hline\hline
		\end{tabular}
		\caption{Special case ($\alpha=0$). Eigenvalues of the linearized system and corresponding stability classification. Entries with at least one zero eigenvalue are non-hyperbolic; in those cases the local behavior is assessed numerically (num.). Ghost and Laplacian stability condition are also included.}
		\label{tab-4}
	\end{table*}

	The case $\alpha=0$ (purely quadratic coupling) exhibits a phase space structure with matter-dominated big-bang-type saddles, de Sitter attractors with SEC violation, stiff-fluid regimes and a continuous family of de Sitter-type points. The quadratic interaction is sufficient to support viable late-time acceleration, with energy conditions and stability properties replicating those found in the more general case. The explicit comparison with the general model ($\alpha\neq0$) shows that the qualitative cosmological behavior is robust against the removal of the linear term. This highlights the universality of accelerated expansion and the structural stability of the model under variations of the coupling parameters.
	
	\subsection{Case $\beta=0$}
	In the absence of the quadratic coupling ($\beta = 0$), the model reduces to $f(R,T^\phi) = R + \alpha T^\phi$, keeping only the linear trace interaction. This limit enables a focused analysis of the physical role of the parameter $\alpha$ in driving the cosmological dynamics. The fact that $\beta=0$ fixes $z=1$, reducing the dimension of the phase space. In this scenario the critical points read: 
	
	\begin{itemize}
		\item Points $r_1^\pm = (\pm 1, 1, 1)$:
		These represent matter-dominated saddle points with $\Omega_m = 1$, undetermined $\omega_\phi$ and a decelerated expansion $q = 1/2$. The eigenvalues of the linearized system are $\{-3, -\tfrac{3}{2}, \tfrac{3}{2}\}$, characterizing a saddle node. The solution is free of Laplacian instabilities ($c_s^2=1$), while the scalar sector is degenerate  ($Q_s=0$). All the energy conditions are saturated. These points generalize the classic matter-dominated era in GR and are always present, independent of $\alpha$.
		
		\item Points $r_2^\pm = (\pm 1, 1/2, 1)$:
		These are scalar field-dominated de Sitter attractors, existing only for $\alpha=0$. The universe is fully dominated by the scalar field $\Omega_m = 0$, with $\omega_\phi = -1$ and $q = -1$. The eigenvalues are depicted by $\{-3, -3, 0\}$, indicating a non-hyperbolic attractor. The solution is free of Laplacian instabilities ($c_s^2=1$), while the scalar sector is degenerate  ($Q_s=0$). Here, NEC and DEC are saturated, WEC is satisfied and SEC is violated. These points show that, in the absence of the linear coupling, late-time acceleration is still possible, matching the behavior found in the purely quadratic and general cases.
		
		\item Points $r_3^\pm = (\pm 1/2, 1, 1)$:
		Stiff-fluid saddle points, existing only for vanishing $\alpha$. The universe is dominated by the kinetic energy of the scalar field $\Omega_m = 0$, with $\omega_\phi = 1$ and a decelerated expansion $q = 2$. The eigenvalues $\{-6, 3, 3\}$, corresponding to saddle behavior. The solution is free of Laplacian instabilities ($c_s^2=1$), while ($Q_s=3$). Here, NEC, WEC and SEC are satisfied, while DEC is saturated. 
		
		\item Points $r_4^\pm = \left( \frac{1 + \alpha \mp \sqrt{1 + \alpha}}{\alpha}, 1, 1 \right)$:
		These stiff-like solution exist only for $\alpha\neq0$. The universe is in a vacuum matter decelerated phase $q=2$, $\omega_\phi=1$, $\Omega_m=0$ and the eigenvalues are $\{3, 3, 3\}$, indicating a unstable node. Ghost and Laplacian instabilities are absent $Q_s=3$ and $(c_s^2=1)$. As in the previous case, NEC, WEC and SEC are satisfied, while DEC is saturated.
	\end{itemize}
	
	\begin{table*}[htbp]
		\centering
		\begin{tabular}{||c||c|c|c|c|c|c|c|c|c|c|c|c||}
			\hline\hline
			Crit. Point & $x_\pm$ & $y$ & $u$ & Existence & $\Omega_m$ & $q$ & $\omega_\phi$ & NEC & WEC & DEC & SEC \\
			\hline\hline
			$r_1^\pm$ & $\pm 1$ & $1$ & $1$ & always & 1 & 1/2 & und. & saturated & saturated & saturated & saturated \\
			$r_2^\pm$ & $\pm 1$ & $1/2$ & $1$ & $\alpha=0$ & 0 & $-1$ & $-1$ & saturated & satisfied & saturated & violated \\
			$r_3^\pm$ & $\pm 1/2$ & $1$ & $1$ & $\alpha=0$ & 0 & $2$ & $1$ & satisfied & satisfied & saturated & satisfied \\
			$r_4^\pm$ & $\frac{1+\alpha \mp \sqrt{1+\alpha}}{\alpha}$ & $1$ & $1$ & $\alpha\neq0$ & 0 & $2$ & $1$ & satisfied & satisfied & saturated & satisfied \\
			\hline\hline
		\end{tabular}
		\caption{Physically meaningful critical points of the autonomous system for the case $ \beta = 0 $, including cosmological parameters and energy conditions.}
		\label{tab-5}
	\end{table*}
	
	\begin{table*}[htbp]
		\centering
		\begin{tabular}{||c||c|c|c|c|c|c||}
			\hline\hline
			Crit. Point & $\lambda_1$ & $\lambda_2$ & $\lambda_3$ & Stability & $Q_s$ & $c_s^2$ \\
			\hline\hline
			$r_1^\pm$ & $-3$ & $-3/2$ & $3/2$ & saddle & $0$ & $1$ \\
			$r_2^\pm$ & $-3$ & $-3$ & $0$  & non-hyperbolic attractor (num. inv.) & $0$ & $1$ \\
			$r_3^\pm$ & $-6$ & $3$ & $3$ & saddle & $3$ & $1$ \\
			$r_4^\pm$ & $3$ & $3$ & $3$ & unstable node & $3$ & $1$ \\
			\hline\hline
		\end{tabular}
		\caption{Stability of the fixed points for the $ \beta = 0 $ case. Eigenvalues of the linearized system and corresponding stability classification. Entries with at least one zero eigenvalue are non-hyperbolic; in those cases the local behavior is assessed numerically (num.). Ghost and Laplacian stability conditions are also included}
		\label{tab-6}
	\end{table*}
	
	With pure linear coupling $(\beta=0)$, the phase-space contains matter-dominated saddle points and stiff-fluid, decelerated solutions. A de Sitter fixed point arisess only in the special subcase $\alpha=0$, \emph{i.e.}\ for a vanishing linear trace coupling, so accelerated expansion is not generically guaranteed by the purely linear interaction. Consecuently, in this sector the linear coupling by itself does not reproduce all late-time accelerating phases found in the general scenario and the viability of accelerated cosmological evolution requires either the quadratic contribution $(\beta\neq 0)$ or additional model ingredients.

	\subsection{$\alpha= \beta=0$}
	This scenario corresponds to standard quintessence with a minimally coupled scalar field. The gravitational sector reduces to Einstein gravity and the dynamics is entirely governed by the kinetic and potential energy of the scalar field. The trace-dependent terms completely vanish and the system recovers the standard behavior expected in GR with a canonical scalar field. We analyze the behavior of system through the density parameter $ \Omega_m $, the scalar field equation of state $ \omega_\phi$ and the deceleration parameter $ q $. In addition, we evaluate the conditions for ghost and Laplacian instabilities $ Q_s $ and $ c_s^2 $, as well as the standard energy conditions.
	
	\begin{itemize}
		\item Points $o_1^\pm = (\pm 1, 1, 1)$: Decelerated matter-dominated points corresponding to $m \ll H$ and $q = \tfrac{1}{2}$. The matter density entirely dominates with $\Omega_m = 1$ and undetermined equation of state $\omega_\phi$ is presented. The eigenvalues of the linearized system are $\{-3, -\tfrac{3}{2}, \tfrac{3}{2}\}$, indicating a saddle point. The solution is free of Laplacian instabilities ($c_s^2=1$), while the scalar sector is degenerate  ($Q_s=0$). All the energy conditions are saturated.
		
		\item Points $o_2^\pm = (\pm 1, 1/2, 1)$: These critical points are de Sitter attractors characterized by an accelerated expansion $q = -1$, $\omega_\phi = -1$ and complete scalar field domination ($\Omega_m = 0$). The eigenvalues are: $\{-3, -3, 0\}$, and by numerical evaluation we find the stable nature. Here, NEC and DEC are saturated, WEC is satisfied and SEC is violated.The solution is free of Laplacian instabilities ($c_s^2=1$), while the scalar sector is degenerate  ($Q_s=0$).
		
		\item Points $o_3^\pm = (\pm 1/2, 1, 1)$: These correspond to an unstable stiff-fluid points ($\omega_\phi = 1$, $q = 2$), where the universe is fully dominated by the kinetic energy of the scalar field ($\Omega_m = 0$). The stability analysis indicates a unstable node, since $\{3,3,3\}$, with no ghost or Laplacian stability $Q_s = 3$, $c_s^2 = 1$. The energy contidions NEC, WEC and SEC are satisfied and DEC is saturated.
	\end{itemize}
	
	\begin{table*}[htbp]
		\centering
		\begin{tabular}{||c||c|c|c|c|c|c|c|c|c|c|c|c||}
			\hline\hline
			Crit. Point & $x_\pm$ & $y$ & $u$ & Existence & $\Omega_m$ & $q$ & $\omega_\phi$ & NEC & WEC & DEC & SEC \\
			\hline\hline
			$o_1^\pm$ & $\pm 1$ & $1$ & $1$ & always & 1 & 1/2 & und. & saturated & saturated & saturated & saturated \\
			$o_2^\pm$ & $\pm 1$ & $1/2$ & $1$ & always & 0 & $-1$ & $-1$ & saturated & satisfied & saturated & violated \\
			$o_3^\pm$ & $\pm 1/2$ & $1$ & $1$ & always & 0 & $2$ & $1$ & satisfied & satisfied & saturated & satisfied \\
			\hline\hline
		\end{tabular}
		\caption{Physically meaningful critical points of the autonomous system for the case $ \alpha = \beta = 0 $, including cosmological parameters and energy conditions.}
		\label{tab-7}
	\end{table*}
	
	\begin{table*}[htbp]
		\centering
		\begin{tabular}{||c||c|c|c|c|c|c||}
			\hline\hline
			Crit. Point & $\lambda_1$ & $\lambda_2$ & $\lambda_3$ &  Stability & $Q_s$ & $c_s^2$ \\
			\hline\hline
			$o_1^\pm$ & $-3$ & $-3/2$ & $3/2$  & saddle & 0& $1$ \\
			$o_2^\pm$ & $-3$ & $-3$ & $0$  & non-hyperbolic attractor & 0 & $1$ \\
			$o_3^\pm$ & $3$ & $3$ & $3$ & unstable node & 3 & $1$ \\
			\hline\hline
		\end{tabular}
		\caption{Stability of the fixed points for the $ \alpha = \beta = 0 $ scenario. Eigenvalues of the linearized system and corresponding stability classification. Entries with at least one zero eigenvalue are non-hyperbolic; in those cases the local behavior is assessed numerically (num.). Ghost and Laplacian stability conditions are also included.}
		\label{tab-8}
	\end{table*}
	
	The standard quintessence limit scenario ($\alpha = \beta = 0$) naturally recovers all the familiar cosmological phases of GR with a canonical scalar field. The phase space structure consists of big-bang-type matter-dominated saddle points, stiff-fluid saddle points and a late-time de Sitter attractor. Notably, the violation of the strong energy condition occurs only at the de Sitter point, highlighting the expected behavior for scalar fields with dominant kinetic energy. All solutions in this regime remain free of ghost and Laplacian instabilities, confirming the consistency of the canonical case as a particular limit of the more general $f(R,T^\phi)$ scenario.

	\section{Discussion}\label{discussion}
	In this work, we have explored the cosmological dynamics of the $f(R,T^\phi)$ gravity model with both linear and quadratic trace couplings, $f(R,T^\phi) = R + \alpha T^\phi + \beta (T^\phi)^2/2$. Our dynamical system analysis reveals a remarkably rich phase-space structure reproducing the standard cosmological eras of General Relativity while admitting novel solutions that arise solely due to the nonminimal trace couplings. 
	
	The general scenario ($\alpha \neq 0$, $\beta \neq 0$) is characterized by a spectrum of critical points that can be grouped into matter-dominated saddles, stiff-fluid solutions and a de Sitter and quasi de Sitter family of critical points. Points $p_1^\pm$ and $p_2^\pm$ represent decelerated matter-dominated saddle configurations, with $\Omega_m=1$ and $q=\tfrac{1}{2}$, so they play the role of transient epochs that can connect early-time evolution with the late-time accelerated solutions. The distinction is that $p_1^\pm=(\pm 1,1,0,1)$ lies on the boundary $z=0$ (where the quadratic trace coupling dominates), whereas $p_2^\pm=(\pm 1,1,1,1)$ is in the weak-coupling regime $z\simeq 1$. At both points the scalar contribution vanishes ($\rho_\phi=p_\phi=0$), hence $\omega_\phi$ is undefined. Both scenarios are free from Laplacian instabilities $c_s^2=1$, and the scalar sector is degenerate ($Q_s=0$), so the no-ghost condition is inconclusive at linear order. Points $p_3^\pm$ correspond to scalar-field-dominated de Sitter solutions with $\Omega_m=0$, $\omega_\phi=-1$ and $q=-1$. Their linear spectrum contains two zero eigenvalues and numerical inspection indicates a non-hyperbolic attractor behavior. They are Laplacian-stable ($c_s^2=1$) but degenerate in the scalar sector ($Q_s=0$); NEC and DEC are saturated, WEC is satisfied and SEC is violated, consistently with late-time acceleration. Points $p_4^\pm$ correspond to a one-parameter family of scalar-field-dominated de Sitter--type solutions, with $\Omega_m=0$, $\omega_\phi=-1$ and $q=-1$. They are non-hyperbolic saddles, since $f_1(\alpha,u) \ge 0$ and $f_2(\alpha,u) \le 0$ for $\alpha>0$. The solution is free of Laplacian instabilities ($c_s^2=1$), while the scalar sector is degenerate ($Q_s=0$). Here, NEC and DEC are saturated, WEC is satisfied, and SEC is violated. The quasi de Sitter like solution $p_5^\pm$ is free from ghost and Laplacian instabilities, since $Q_s= 2\alpha(1+\alpha)$ and $c_s^2=0$, respectively. The analysis of their eigenvalues, supported by numerical integration where necessary, confirms their saddle nature. Points $p_6^\pm$ are three-dimensional unstable manifolds, so they act as strong early-time repellers with $\Omega_m=0$, $\omega_\phi=1$ and $q=2$, hence they describe decelerated kinetic regimes. The eigenvalues $\{-6,3,3,3\}$ show that they are saddle points. Ghost and Laplacian instabilities are absent ($Q_s=3$ and $c_s^2=1$). Here, NEC, WEC and SEC are satisfied, while DEC is saturated.

	When restricting to the purely quadratic regime ($\alpha = 0$, $\beta \neq 0$), the model retains all key cosmological phases: early-time matter-like saddles, stiff-fluid points and a continuous family of de Sitter-type solutions. These results emphasize that the quadratic interaction alone suffices to drive the universe towards late-time acceleration. The comparison with the general case shows that the qualitative phase-space structure remains robust: the removal of the linear coupling does not preclude the existence of background accelerated critical points, but does modulate the detailed location and stability of the critical points.
	
	On the other hand, in the purely linear case ($\beta=0$, $\alpha\neq0$), the phase space contains matter-dominated saddles and stiff-fluid decelerated solutions. Late-time de Sitter acceleration is not a generic prediction of the linear trace coupling, since the de Sitter fixed point is recovered only in the special subcase $\alpha=0$. Thus, in this sector the linear coupling alone does not reproduce the full late-time accelerating behavior found in the general model.
	
	Finally, the General Relativity limit ($\alpha = 0$, $\beta = 0$) recovers all the expected behaviors: matter-dominated saddles, stiff-fluid transient points and a late-time de Sitter attractor driven by the scalar potential. The absence of nonminimal couplings naturally eliminates the new families of solutions present in the broader model.
	
	A point of particular interest is the universal appearance of stiff-fluid points in all scenarios. While these solutions are mathematically consistent and free from instabilities, their cosmological relevance is restricted to very early times: observational constraints from nucleosynthesis and the cosmic microwave background exclude any prolonged stiff-fluid era in the post-recombination universe. In our analysis, these points always appear as saddle or unstable nodes, ensuring that cosmic trajectories naturally evolve towards the more physically acceptable attractors.
	
	A further robust feature is the violation of the strong energy condition at the background level for the accelerated (de Sitter or de Sitter-like) critical solutions, as expected for late-time acceleration. However, several accelerated points belong to a degenerate scalar sector with $Q_s=0$, where the standard linear no-ghost diagnostics are inconclusive, while the non-degenerate accelerated solution features $c_s^2=0$, i.e. a marginal gradient sector. Therefore, a definitive statement about perturbative viability requires an analysis beyond the linear criteria adopted here.
	
	Despite these positive features, certain limitations remain. Our analysis has been restricted to homogeneous and isotropic backgrounds; possible pathologies or instabilities could arise at the level of perturbations, especially in the presence of more general scalar potentials or additional matter sectors. Furthermore, the constraints derived from the dynamical analysis must be supplemented with detailed confrontation with cosmological observations, particularly concerning the growth of structure and the CMB.

	
	\section{Conclusions}\label{conclusions}
	The dynamical analysis presented in this work shows that the model $f(R,T^\phi)=R+\alpha T^\phi+\beta (T^\phi)^2/2$ exhibits a substantially richer phase-space structure than the minimally coupled limit. By recasting the cosmological equations into a compact autonomous system, we identified isolated critical points and continuous families describing matter-dominated epochs, stiff-fluid-like kinetic regimes, scalar-field-dominated solutions and accelerated (de Sitter-like) expansion at the background level. These features arise from the non-minimal trace couplings and disappear in the limit $\alpha=\beta=0$, where the dynamics reduce to standard minimally coupled quintessence.
	
	Our results indicate that robust late-time acceleration is primarily associated with the quadratic trace sector: the $(T^\phi)^2$ contribution admits accelerated de Sitter-like critical solutions at the background level and generates continuous families of critical points absent in the canonical case. By contrast, in the purely linear sector ($\beta=0$, $\alpha\neq0$) late-time de Sitter acceleration is not a generic outcome, since the de Sitter fixed point is recovered only in the special subcase $\alpha=0$.
	
	We also examined energy conditions and the perturbative behavior using the standard no-ghost and gradient criteria. At the background level, accelerated critical solutions violate the strong energy condition as expected, while the null and weak energy conditions can remain satisfied or saturated depending on the branch. However, several accelerated points lie in a degenerate scalar sector with $Q_s=0$, where the usual linear perturbation diagnostics are inconclusive, whereas the non-degenerate accelerated solution features $c_s^2=0$, i.e. a marginal gradient sector. Therefore, establishing full perturbative viability requires an analysis beyond the linear treatment adopted here.
	
	Broadly speaking, trace-coupled $f(R,T^\phi)$ cosmologies provide a natural extension of scalar-field models and can reproduce the standard cosmological epochs while allowing additional accelerated solutions sourced by geometric-matter interactions. A dedicated phenomenological study including observational constraints and a perturbation analysis beyond the degenerate regime is needed to assess whether this framework can serve as a viable alternative to standard DE scenarios.

	\acknowledgments
	
	The authors are grateful to SNII-CONACyT for continuous support of their research activity and to FORDECYT-PRONACES-CONACYT for support of the present research under grant CF-MG-2558591. We appreciate the facilities provided by the Universidad Michoacana de San Nicolás de Hidalgo and the CIC-UMSNH during the realization of this investigation.
	

\end{document}